\pdfoutput=1
\documentclass[cernpreprint,UKenglish,texmf]{atlasdoc}
 
\usepackage{multirow}
\usepackage{mathtools}
\usepackage[subfigure=true]{atlaspackage}
\usepackage{atlaspackage}
\usepackage{atlasbiblatex}
 
\usepackage{atlasphysics}
 
\graphicspath{{logos/}{figures/}}
 
\usepackage{ANA-HDBS-2018-18-PAPER-defs}

\AtlasTitle{Search for the $HH \rightarrow b \bar{b} b \bar{b}$ process via vector-boson fusion production using proton--proton collisions at $\sqrt{s} = 13$~\TeV\ with the ATLAS detector}

\AtlasAbstract{
A search for Higgs boson pair production via vector-boson fusion (VBF) in the $b\bar{b}b\bar{b}$ final state is carried out with the ATLAS experiment using 126~fb$^{-1}$
of proton--proton collision data delivered at $\sqrt{s} = 13$~\TeV\ by the Large Hadron Collider. This search is sensitive to VBF production of additional heavy bosons that
may decay into Higgs boson pairs, and in a non-resonant topology it can constrain the quartic coupling between the Higgs bosons and vector bosons. No significant excess
relative to the Standard Model expectation is observed, and limits on the production cross-section are set at the 95\% confidence level for a heavy scalar
resonance in the context of an extended Higgs sector, and for non-resonant Higgs boson pair production. Interpretation in terms of the coupling between a Higgs boson
pair and two vector bosons is also provided: coupling values normalised to the Standard Model expectation of $\kappa_{2V} < -0.43$ and $\kappa_{2V} > 2.56$ are excluded
at the 95\% confidence level in data.
}
 
\author{The ATLAS Collaboration}
 
\AtlasRefCode{HDBS-2018-18}
 
\PreprintIdNumber{CERN-EP-2019-267}

\AtlasJournalRef{JHEP 07 (2020) 108}
\AtlasDOI{10.1007/JHEP07(2020)108}
 
\hypersetup{pdftitle={ATLAS document},pdfauthor={The ATLAS Collaboration}}
 
\begin{document}
 
\maketitle

 
\section{Introduction}
 
The Higgs boson ($H$) was discovered by the ATLAS and CMS collaborations in 2012~\cite{HIGG-2012-27,CMS-HIG-12-028} using proton--proton ($pp$) collisions at the Large Hadron Collider (LHC). The measured properties have so far been found to be in agreement with the Standard Model (SM)
predictions. The production of a pair of Higgs bosons ($HH$) is a rare process in the SM with a cross-section about 1000 times smaller than the single Higgs boson production cross-section, but various theories beyond the SM (BSM) predict cross-sections for $HH$ production that are significantly higher than the SM prediction. Spin-0 resonances, with narrow or broad width, that decay into Higgs boson pairs, appear in BSM scenarios~\cite{PhysRevD.8.1226,Branco:2011iw}. Enhanced non-resonant Higgs boson pair production is predicted by many models, for example those featuring light coloured scalars \cite{Kribs:2012kz} or new contact interactions, such as direct $t\bar{t}HH$ vertices~\cite{Grober:2010yv,Contino:2012xk}.

Previous searches for Higgs boson pair production in the $b\bar{b}b\bar{b}$ channel were carried out in the gluon--gluon fusion (ggF) production mode by the ATLAS and CMS collaborations~\cite{EXOT-2016-31,CMS-HIG-17-009,CMS-B2G-16-026,CMS-B2G-17-019,CMS-HIG-17-017}, and limits were set for resonant and non-resonant production. Statistical combinations of search results for $HH$ in various decay channels were also performed by the two experiments~\cite{HDBS-2018-58,Sirunyan:2018two}, profiting from the sensitivity of several final states.

\begin{figure*}[b]
\begin{center}
\subfigure[]{\includegraphics[width=0.24\hsize]{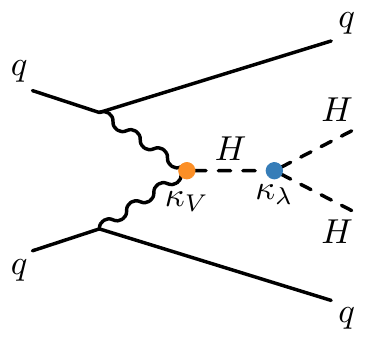}\label{fig:feyndiagrams:cvlambda}}
\subfigure[]{\includegraphics[width=0.24\hsize]{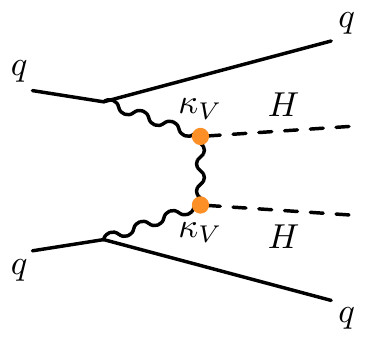}\label{fig:feyndiagrams:cvcv}}
\subfigure[]{\includegraphics[width=0.24\hsize]{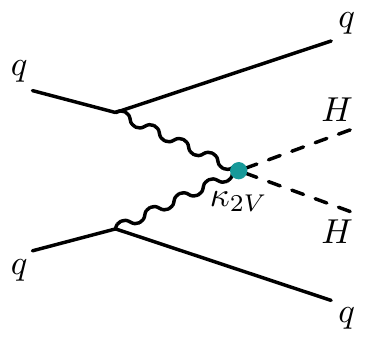}\label{fig:feyndiagrams:c2v}}
\subfigure[]{\includegraphics[width=0.24\hsize]{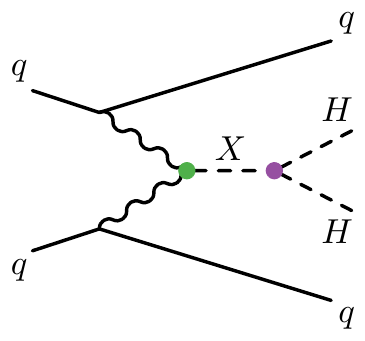}\label{fig:feyndiagrams:res}}
\caption{Tree-level Feynman diagrams contributing to Higgs boson pair production via VBF. Diagrams~\subref{fig:feyndiagrams:cvlambda}, \subref{fig:feyndiagrams:cvcv} and \subref{fig:feyndiagrams:c2v}
illustrate the non-resonant production modes scaling with $\kv\kappa_{\lambda}$, $\kappa^2_V$ and \ktv, respectively. Diagram~\subref{fig:feyndiagrams:res} illustrates the resonant production mode.}
\label{fig:feyndiagrams}
\end{center}
\end{figure*}
 
This paper focuses on searches for Higgs boson pair production via vector-boson fusion (VBF), through diagrams such as those presented in Figure~\ref{fig:feyndiagrams}, and using the dominant $H \rightarrow b\bar{b}$ decay mode~\cite{Bishara:2016kjn}. The VBF process (\vbfproc) is characterised by the presence of two jets ($j$) with a large rapidity gap resulting from quarks from which a vector boson ($V$) is radiated. In the SM, three different types of couplings are involved in $HH$ production via VBF: the Higgs boson self-coupling ($HHH$), the  Higgs-boson--vector-boson coupling ($VVH$) and the quartic (di-vector-boson--di-Higgs-boson, or $VVHH$) coupling. The coupling modifiers $\kappa_\lambda$, \kv\ and \ktv\ control the strength of the $HHH$, $VVH$ and $VVHH$ couplings with respect to the SM value, respectively, and are normalised so that they are equal 1 in the SM. A deviation of these coupling modifiers from their SM expectations could lead to enhanced $HH$ production. While searches in the ggF mode are more sensitive to deviations in $\kappa_\lambda$, the VBF topology has unique sensitivity to \ktv~\cite{Bishara:2016kjn} because the ggF mode does not involve the $VVHH$ interaction. For resonant production, two classes of signals are tested to perform a generic inclusive search for resonances with masses \mscalar\ in the range \mlow--\mhigh~\GeV. The first signal class is representative of a broad resonance with width typically 10-20\% of the signal mass; it corresponds to a heavy scalar of the 2HDM Type II model~\cite{Coleppa:2013dya} and is obtained by setting the ratio of vacuum expectation values of the two Higgs doublets $\tan(\beta) = 2.0$ and $\sin(\beta-\alpha) = 0.6$, where $\alpha$ is the mixing angle between the two CP-even Higgs bosons. The second class features a narrow resonance with a fixed width of 4~\MeV.
 
\newcommand{\AtlasCoordFootnote}{
ATLAS uses a right-handed coordinate system with its origin at the nominal interaction point (IP)
in the centre of the detector and the \(z\)-axis along the beam pipe. The \(x\)-axis points from the IP
to the centre of the LHC ring, and the \(y\)-axis points upwards. Cylindrical coordinates \((r,\phi)\) are
used in the transverse plane, \(\phi\) being the azimuthal angle around the \(z\)-axis.
The pseudorapidity is defined in terms of the polar angle \(\theta\) as \(\eta = -\ln \tan(\theta/2)\).
Angular distance is measured in units of \(\Delta R \equiv \sqrt{(\Delta\eta)^{2} + (\Delta\phi)^{2}}\).}

\section{ATLAS detector}
 
The ATLAS experiment~\cite{PERF-2007-01,ATLAS-TDR-2010-19,PIX-2018-001} at the LHC operates a multipurpose particle detector with a
forward--backward symmetric cylindrical geometry and a near \(4\pi\) coverage in solid angle.\footnote{\AtlasCoordFootnote}
It consists of an inner tracking detector surrounded by a thin superconducting solenoid providing a \SI{2}{\tesla}
axial magnetic field, electromagnetic and hadronic calorimeters, and a muon spectrometer. The inner tracking detector
covers the pseudorapidity range \(|\eta| < 2.5\). It consists of silicon pixel, silicon microstrip, and transition-radiation
tracking detectors. Lead/liquid-argon (LAr) sampling calorimeters provide electromagnetic (EM) energy measurements
with high granularity. A hadronic steel/scintillator-tile calorimeter covers the central pseudorapidity range (\(|\eta| < 1.7\)).
The endcap and forward regions are instrumented with LAr calorimeters for both the EM and hadronic energy measurements
up to \(|\eta| = 4.9\). The muon spectrometer surrounds the calorimeters and incorporates three large air-core toroidal superconducting
magnets with eight coils each. The field integral of the toroids ranges between \num{2.0} and \SI{6.0}{Tm}
across most of the detector. The muon spectrometer includes a system of precision tracking chambers and fast detectors for triggering~\cite{TRIG-2016-01}.
A two-level trigger system is used to select events. The first-level trigger is implemented in hardware and uses a subset of the detector
information to reduce the accepted rate to at most \SI{100}{\kilo\hertz}. This is followed by the software-based high-level
trigger (HLT), which reduces the accepted event rate to \SI{1}{\kilo\hertz} on average.
 
 
\section{Data and simulated samples}
\label{sec:samples}

This search is performed using data collected by the ATLAS experiment between 2016 and 2018 in $\sqrt{s}=13$~\TeV\ LHC $pp$ collisions, which correspond
to an integrated luminosity of \lumi~fb$^{-1}$. Only events recorded during stable beam conditions and when the detector components relevant to the analysis
were operating normally are considered~\cite{DAPR-2018-01}. During the 2016 data-taking, a fraction of the data was affected by an inefficiency in the vertex reconstruction in the HLT,
which reduced the efficiency of the algorithms used to identify jets originating from $b$-hadron decays; those events were not retained for further analysis. This
reduces the integrated luminosity of the 2016 dataset to 24.3~fb$^{-1}$.

Simulated Monte Carlo (MC) event samples are used to model signal production and the backgrounds from top-quark-pair (\ttbar) production.  The dominant process
arises from multijet production, and is modelled using data-driven techniques.
 
Events with a generic scalar resonance produced via VBF and decaying into $HH \to$ \bbbb\ were generated with \textsc{Powheg-Box}~v2~\cite{Nason:2004rx,Frixione:2007vw,Alioli:2010xd}
interfaced to \PYTHIA~8.186~\cite{Sjostrand:2007gs} for parton showering and hadronisation, with the Higgs boson mass fixed to 125~\gev~\cite{HIGG-2014-14}. The NNPDF23LO
parton distribution function (PDF) set~\cite{Ball:2012cx} and the A14 set of tuned parameters~\cite{ATL-PHYS-PUB-2014-021} for
underlying-event simulation were used. Resonant signal samples were produced with masses ranging from 260~\gev\ to 1000~\gev.
 
Non-resonant production of Higgs boson pairs was simulated with \textsc{MadGraph5\_aMC@NLO}~\cite{Alwall:2014hca}. The Higgs boson self-coupling and the couplings of the Higgs boson to vector bosons were set to their SM values, while the \ktv\ coupling modifier was varied. Samples with \ktv\ equal to 0.0, 0.5, 1.0, 1.5, 2.0, and 4.0 were generated at leading order (LO). Interference between various diagrams contributing to the non-resonant signal is considered in the simulation. A linear combination of three samples is used to derive distributions for a finer granularity of \ktv\ values, following a technique used to generate \kl\ distributions~\cite{HDBS-2018-58}.  A particular choice of three samples is done based on the \ktv\ range of the generated distributions to avoid large weights and to reduce statistical uncertainties. The cross-section of the VBF $HH$ process, evaluated at next-to-next-to-next-to-leading order (N$^3$LO) in QCD, is $1.73\pm 0.04$~fb in the SM~{\cite{Baglio:2012np,Liu-Sheng:2014gxa,Dreyer:2018qbw,Dreyer:2018rfu}. The N$^3$LO to LO cross-section ratio at the SM value is calculated and this factor is applied to the cross-sections at each \ktv\ point.

To estimate the contribution from ggF $HH$ production with two additional jets that can mimic the VBF $HH$ topology, the SM non-resonant production of Higgs boson pairs via ggF was simulated with \textsc{MadGraph5\_aMC@NLO} using the CT10 PDF set~\cite{Lai:2010vv} and the FTapprox method~\cite{Grazzini:2018bsd} to include finite top-quark mass effects. In this sample, the generation of $pp \rightarrow H + \text{parton}$ is done at next-to-leading order (NLO). Parton showers and hadronisation were simulated with Herwig 7.0.4~\cite{Bellm:2015jjp}. Interference effects with other SM processes are found to be marginal and are ignored. The cross-section is evaluated at next-to-next-to-leading order (NNLO) with the resummation at next-to-next-to-leading-logarithm (NNLL) accuracy and including top-quark mass effects at NLO~\cite{Dawson:1998py,Borowka:2016ehy,Baglio:2018lrj,deFlorian:2013jea,Shao:2013bz,deFlorian:2015moa,Grazzini:2018bsd}; it is equal to $31.05^{+1.40}_{-1.99}$~fb. The uncertainty includes the variations of the factorisation and renormalisation scales, PDF and \alphas.
 
The generation of \ttbar\ events was performed with \textsc{Powheg-Box}~v2~\cite{Frixione:2007nw} using the NNPDF3.0NLO~\cite{Ball:2014uwa}  PDF set. The parton showers,
hadronisation, and underlying event were simulated using \PYTHIA~8.230~\cite{Sjostrand:2014zea} with the NNPDF23LO PDF set and the corresponding A14 set of tuned underlying-event parameters. The predicted $t\bar{t}$ production cross-section is $831.8^{+19.8}_{-29.2} \pm 35.1$~pb as calculated with the Top++~2.0 program to NNLO in perturbative QCD, including soft-gluon resummation to NNLL accuracy~\cite{Czakon:20142930}, and assuming a top-quark mass of 172.5~\gev. The first uncertainty comes from the independent variations of the factorisation and renormalisation scales, while the second one is associated with variations in the PDF and \alphas, following the PDF4LHC prescription with the MSTW2008 68\% CL NNLO, CT10 NNLO and NNPDF2.3 5f FFN PDF sets~\cite{Botje:2011sn,Martin:2009bu,Gao:2013xoa,Ball:2012cx}.

For all simulated events, $c$-hadron and $b$-hadron decays were handled by {\textsc{EvtGen}}~1.2.0~\cite{Lange:2001uf}. To simulate the impact of
multiple $pp$ interactions that occur within the same or nearby bunch crossings (pile-up), minimum-bias events generated with
\PYTHIA~8.186 using the NNPDF2.3LO set of PDFs and the A3 set of tuned parameters~\cite{ATL-PHYS-PUB-2016-017} were overlaid on the hard-scatter
process. The detector response was simulated with \GEANT~4~\cite{Agostinelli:2002hh, SOFT-2010-01}, and the events were processed with the same reconstruction software as that was used for the data.
 
 
\section{Event reconstruction}
\label{sec:reconstruction}
 
Events are required to have at least one reconstructed primary vertex with at least two associated tracks, each with transverse momentum $\pT > 0.4$~\gev. For events
with more than one primary-vertex candidate, the one with the largest track $\sum\pT^2$ is chosen as the hard-scatter primary vertex.
 
Jets are reconstructed from three-dimensional topological clusters of energy deposits in the calorimeter~\cite{PERF-2014-07} with the anti-$k_t$ algorithm~\cite{Cacciari:2008gp}
implemented in the \textsc{FastJet} package~\cite{Fastjet} with radius parameter $R = 0.4$. Clusters are calibrated at the EM scale~\cite{PERF-2016-04} and their
energy is corrected for additional energy deposits from pile-up interactions using an area-based correction \cite{Cacciari:2008gn}. Subsequently, calibration using
\pt- and $\eta$-dependent factors derived from simulation is applied, followed by the global sequential calibration~\cite{PERF-2016-04}. The latter reduces the flavour
dependence of the calibration and energy leakage effects. The final calibration is based on in situ measurements in collision data~\cite{PERF-2016-04}. To preferentially reject jets originating from pile-up interactions, a multivariate classification algorithm (jet vertex tagger) based on tracking information~\cite{PERF-2014-03} is used for jets with
$\pt < 60$~\GeV\ and $|\eta| < 2.4$. The selected working point provides an inclusive hard-scatter process
efficiency of about 97\% in that kinematic region. The efficiency in the simulation is corrected to match that measured in data. Events having jets consistent with noise in
the calorimeter or non-collision backgrounds are vetoed~\cite{PERF-2016-04}.
 
Jets containing $b$-hadrons are identified using a multivariate algorithm (MV2c10)~\cite{FTAG-2018-01,ATL-PHYS-PUB-2017-013}, which exploits information about
the jet kinematics, the impact parameters of tracks associated with the jet and the presence of displaced vertices to form the decision. The $b$-tagging requirements result
in an efficiency of 70\% for jets with $\pt > 20$~\gev\ containing $b$-hadrons, and the misidentification rate is $0.3\%$ ($11.2\%$) for light-flavour (charm) jets. These
were determined in a sample of simulated \ttbar\ events. For all simulated events the $b$-tagging efficiencies are corrected to match those measured in
data~\cite{FTAG-2018-01,ATLAS-CONF-2018-001,ATLAS-CONF-2018-006}.
 
To further correct the $b$-jet energy for effects that are not considered in the default calibration, a jet energy regression is used. The method uses a boosted
decision tree (BDT) algorithm implemented in TMVA~\cite{TMVA}. The BDT training is performed using variables the $b$-jet energy resolution is sensitive to: MV2c10 score, energy leakage outside the jet cone, pile-up contamination, hard radiation from the original parton, and energy loss through
semileptonic $b$-hadron decays. Both the training and the validation are performed with simulated \ttbar\ samples, resulting in approximately 10\% improvement
in the jet energy resolution. The performance of the jet energy regression is validated in $Z(\rightarrow \mu\mu)+b/b\bar{b}$ events in data and no mismodelling is found. The \Hbb\ mass peak is found to be closer to 125~\gev\ and the standard deviation divided by the mean of the mass distribution
is improved by about 25\% for the $m_X = 600$~\gev\ signal sample.

 
 
\section{Event selection}
\label{sec:selection}
 
Events are selected using a combination of $b$-jet triggers, with the lowest jet transverse energy, \ET, threshold at 35~\gev, jet  $|\eta| < 2.5$ and one or two $b$-tagged jets. The $b$-jet trigger efficiency is measured in data, and the simulated events are corrected to match the measured trigger efficiency.
 
To select events compatible with VBF production of Higgs boson pairs decaying into four $b$-quarks, exactly four central $b$-tagged jets with
$\pt > 40$~\GeV\ and $|\eta| < 2.0$ and at least two forward jets with $\pt > 30$~\gev\ and $|\eta| > 2.0$ are required. Events with more than four central jets are
vetoed in case more than four jets fulfil the $b$-tagging requirement. To form Higgs boson
candidates, $b$-tagged jets are used, and the forward jets are considered
as possible remnants of the VBF process. The Higgs boson reconstruction procedure is the same as the one described in Ref.~\cite{EXOT-2016-31}, except that
the usage of the jet energy regression changes the numerical values in the signal region definition described below. A summary of the selection criteria is provided in
Table~\ref{table:selections}.
 
\begin{table}
\centering
\small
\caption{Summary of the selection criteria for capturing the VBF topology, identifying $HH \to b \bar{b} b \bar{b}$ decays, and suppressing background events. Possible remnants of the VBF process are identified using the two highest-\pt forward jets. Labels ``lead'' and ``subl'' refer to the leading and subleading Higgs boson candidates (ordered in \pt), respectively. The definitions of the different analysis regions are also provided. The transverse momenta and masses are expressed in \GeV.}
\label{table:selections}
\begin{tabular}{|c||c|c|}
\hline
& \multicolumn{2}{c|}{Selections}\\
\hline
\hline
\multirow{2}{*}{VBF topology}  & \multirow{2}{*}{\shortstack[c]{At least two jets \\ with \pt > 30, $|\eta| > $2.0 }} & Two highest-\pt jets with opposite sign $\eta$ \\
\cline{3-3}
&         &\detavbf > 5.0 and \mjjvbf > 1000 \\
\hline
\hline
\multirow{8}{*}{Signal topology} & \multicolumn{2}{c|}{Exactly 4 $b$-tagged jets with \pt > 40, $|\eta| < $2.0} \\
\cline{2-3}
& \multirow{2}{*}{If $\mfb <$ 1250} &$\frac{360}{\mfb} - 0.5 < \drbb^{\mathrm{lead}} < \frac{653}{\mfb} + 0.475$\\
&                                      & $\frac{235}{\mfb}\hphantom{- 0.05}  <\drbb^{\mathrm{subl}} < \frac{875}{\mfb} + 0.35\hphantom{0} $\\
\cline{2-3}
& \multirow{2}{*}{If $\mfb \geq$ 1250} & $ \drbb^{\mathrm{lead}} < 1$\\
&                                      & $ \drbb^{\mathrm{subl}} < 1$\\
\cline{2-3}
 
& \multicolumn{2}{c|}{Pairs with minimum} \\
& \multicolumn{2}{c|}{$D_{HH} = \sqrt{(m_{2b}^\text{lead})^2+(m_{2b}^\text{subl})^2} \left| \sin \left( \tan^{-1} \left( \frac{m_{2b}^\text{subl}}{m_{2b}^\text{lead}} \right) - \tan^{-1} \left( \frac{116.5}{123.7}\right) \right) \right| $} \\[3.5pt]
 
\hline
\hline
\multirow{5}{*}{Background rejection}& \multirow{4}{*}{Multijet} & $|\detahh| < 1.5$ \\
\cline{3-3}
&                                      & $|\Sigma_i \vec{\pt}_i|$ < 60, where $i$ = $b$-jets and VBF-jets \\
\cline{3-3}
&                                      & $p_{\text{T},H}^\text{lead} > 0.5\mfb - 103$ \\
\cline{3-3}
&                                      & $p_{\text{T},H}^\text{subl} > 0.33\mfb - 73$ \\
\cline{2-3}
&\ttbar  & Veto if $X_{Wt} = \sqrt{\left( \frac{m_W - 80.4}{0.1 m_W} \right)^2 + \left( \frac{m_t - 172.5}{0.1 m_t} \right)^2 } \le1.5$ \\
\hline
\hline
\multirow{3}{*}{Region definition}  & Signal region (SR) & $X_{HH} = \sqrt{ \left(\frac{m_{2b}^\text{lead}-123.7}{11.6}\right)^2 + \left(\frac{m_{2b}^\text{subl}-116.5}{18.1}\right)^2 } < 1.6$ \\
\cline{2-3}
 
& Validation region (veto SR)  & $\sqrt{ \left( m_{2b}^{\mathrm{lead}} - 123.7\right)^2 + \left( m_{2b}^{\mathrm{subl}} - 116.5\right)^2} < 30$ \\

\cline{2-3}
& Sideband region (veto SR, VR)  & $\sqrt{ \left( m_{2b}^{\mathrm{lead}} - 123.7\right)^2 + \left( m_{2b}^{\mathrm{subl}} - 116.5\right)^2} < 45$ \\
\hline
\end{tabular}
\end{table}

\subsection{VBF-jets selection}
 
The two highest-\pt forward jets with opposite sign of $\eta$ are considered as remnants of the VBF production process if the absolute value of the pseudorapidity
separation between them, \detavbf, exceeds 5.0 and their invariant mass, \mjjvbf, is greater than 1000~\gev.

\subsection{Signal kinematics selection}
 
The four central $b$-tagged jets are considered in three possible combinations of two-jet pairings. Their invariant mass, $\mfb$,
is used to define criteria to select signal-like events. The following $\Delta R$ requirements must be satisfied; these
reflect the correlation of $\mfb$ with the Lorentz boost of the Higgs bosons and the angle between their decay products in the laboratory frame:
\begin{eqnarray*}
\begin{drcases}
\frac{360~\GeV}{\mfb} - 0.5 < \drbb^{\mathrm{lead}} < \frac{653~\GeV}{\mfb} + 0.475\\
\frac{235~\GeV}{\mfb}\qquad\ <\drbb^{\mathrm{subl}} < \frac{875~\GeV}{\mfb} + 0.35
\end{drcases}
\mathrm{if}\ \mfb < 1250~\GeV,
\end{eqnarray*}
\begin{eqnarray*}
\begin{drcases}
\drbb^{\mathrm{lead}} < 1\\
\drbb^{\mathrm{subl}} < 1
\end{drcases}
\mathrm{if}\ \mfb \geq 1250~\GeV,
\end{eqnarray*}
where $\drbb^{\mathrm{lead}}$ and $\drbb^{\mathrm{subl}}$ are the angular distances between the jets that form, respectively, the leading and subleading Higgs boson
candidates (ordered in \pT). These criteria are optimised for both non-resonant and resonant Higgs boson pair production, and the numerical values are chosen to maximise
the signal sensitivity.
 
Out of the possible pairings fulfilling the previous selection, the combination that leads to pairs with a dijet mass closest to the SM Higgs boson mass should be the optimal
choice. However, due to energy loss through semileptonic $b$-hadron decays, this criterion is relaxed. The mass values of 123.7~\GeV\ for the leading Higgs boson candidate and
116.5~\GeV\ for the subleading Higgs boson candidate are found to maximise the signal significance for a resonance with a mass of 600~\GeV, which lies in the middle of the
covered mass range. The same target values are used for all other signal hypotheses. For a given pairing, the quantity \Dhh\ that corresponds to the distance of the leading
and subleading Higgs boson candidate masses, in the (\leadm, \sublm) plane, from the line connecting $\left(0~\GeV, 0~\GeV\right)$ and $\left(123.7~\GeV, 116.5~\GeV\right)$,
can be computed as:
\begin{eqnarray*}
D_{HH} = \sqrt{(m_{2b}^\text{lead})^2+(m_{2b}^\text{subl})^2} \left| \sin \left( \tan^{-1} \left( \frac{m_{2b}^\text{subl}}{m_{2b}^\text{lead}} \right) - \tan^{-1} \left( \frac{116.5~\gev}{123.7~\gev}\right) \right) \right|.
\end{eqnarray*}
The pairing with the smallest value of \Dhh\ is chosen. Studies based on simulation indicate that for SM non-resonant $HH$ production the correct pairs are identified in at least 83\% of the signal events, while for broad resonances the corresponding fraction is greater than 91\%.

\subsection{Selection for background suppression}
 
In order to enhance the sensitivity to signal, various requirements are applied to suppress the background. The magnitude of the vector sum of the transverse
momenta of the selected four $b$-jets
and the two VBF-jets tends to peak at lower values for signal events than for multijet events. Consequently, it is required to be less than 60~\gev.
The pseudorapidity difference between the reconstructed Higgs boson candidates, $|\detahh|$, is required to be below 1.5, and mass-dependent requirements
on the transverse momenta of the leading and subleading Higgs boson candidates, respectively \ptlead\ and \ptsubl,  are:
\begin{eqnarray*}
p_{\text{T},H}^\text{lead} > 0.5\mfb - 103~\GeV, \\
p_{\text{T},H}^\text{subl} > 0.33\mfb - 73~\GeV.
\end{eqnarray*}
 
The resulting dijet pairs are still dominated by multijet events. To further increase the search sensitivity, the dijet masses are required to fulfil:
\begin{eqnarray}
X_{HH} = \sqrt{ \left(\frac{m_{2b}^{\mathrm{lead}}-123.7~\mathrm{GeV}}{11.6~\mathrm{GeV}}\right)^2 + \left(\frac{m_{2b}^{\mathrm{subl}}-116.5~\mathrm{GeV}}{18.1~\mathrm{GeV}}\right)^2 } < 1.6,
\label{eqn:xhh}
\end{eqnarray}
where 11.6~\GeV\ and 18.1~\GeV\ are the experimental widths of the simulated leading and subleading Higgs boson candidates, respectively. The mass resolution of the subleading Higgs boson candidate is worse because
it is composed of the lower \pT jets. These values are derived using a 600~\gev\ resonant signal sample and are similar for other signal samples.
 
Additional requirements are imposed to reduce the number of hadronically decaying \ttbar events by vetoing candidate events compatible with a top-quark decay. The jet with
the highest $b$-tagging score is considered as the $b$-jet originating from a top-quark decay and the remaining central jets are considered to stem from the $W$-boson decay.
Since the top quarks are expected to be produced centrally, only central jets are tested. All possible two-jet combinations in the event are tested and the selected
combination is the one with the smallest value of $X_{Wt}$, defined as:
\begin{eqnarray*}
X_{Wt} = \sqrt{\left( \frac{m_W - 80.4~\GeV}{0.1 m_W} \right)^2 + \left( \frac{m_t - 172.5~\GeV}{0.1 m_t} \right)^2 },
\end{eqnarray*}
where $m_W$ and $m_t$ are the reconstructed invariant masses of the $W$-boson and top-quark candidates, respectively. The event is vetoed
if $X_{Wt}  \le 1.5$. This requirement reduces the \ttbar contamination by about 50\% with negligible impact on the signal efficiency.
 
All requirements listed above define the signal region (SR). The number of selected signal events divided by the number of generated events after each selection step (cumulative acceptance times
efficiency) is shown in Figure~\ref{fig:resolvedEfficiency} for the non-resonant signal as a function of the \ktv\
coupling modifier and for the resonant signal models as a function of the generated mass. The acceptance times efficiency increases as a function of the resonance mass,
while for the non-resonant signal a significant drop is observed at $\ktv = 1$.
The trigger and jet selection requirements cause the drop for \ktv\ values around $1$, while the smaller acceptance times efficiency for low-mass resonances is due to the softer \pt\ spectrum of $b$-jets.
 
\begin{figure*}
\begin{center}
\includegraphics[width=\textwidth]{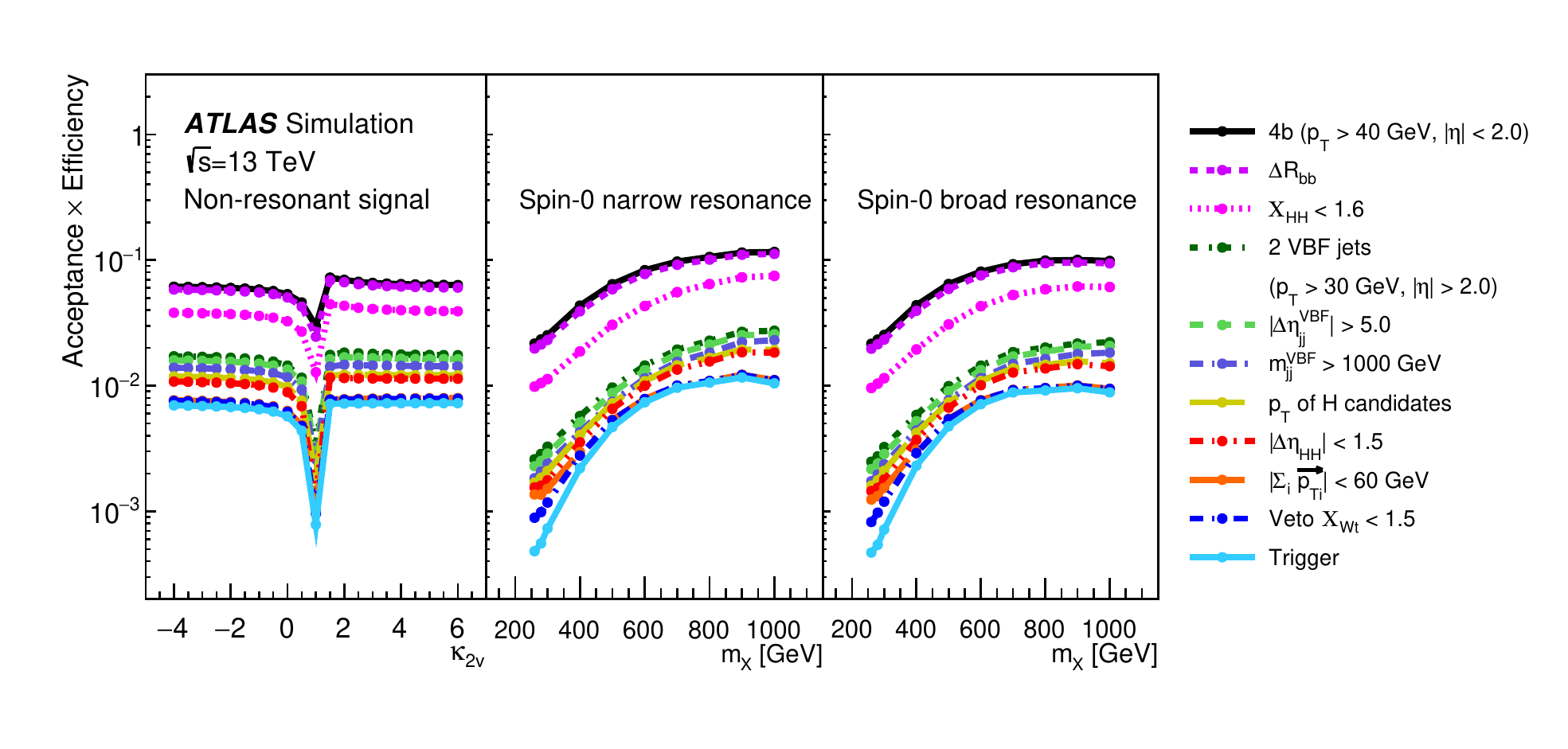}
\caption{Cumulative acceptance times efficiency at each stage of the event selection, as detailed in Section~\ref{sec:selection}. The number of events surviving the selection divided by the number of
generated events is reported separately for the non-resonant signal as a function of the \ktv\ coupling modifier and for the narrow- and broad-width resonance production hypotheses as a function of the generated mass.}
\label{fig:resolvedEfficiency}
\end{center}
\end{figure*}

To estimate the background and to validate the background estimation technique, two regions orthogonal to the SR are used: the
sideband region (SB) and the validation region (VR). The events in the SB and VR must fail the requirement defined in Eq.~(\ref{eqn:xhh})
and fulfil
\begin{eqnarray*}
\sqrt{ \left( m_{2b}^{\mathrm{lead}} - 123.7~\gev\right)^2 + \left( m_{2b}^{\mathrm{subl}} - 116.5~\gev\right)^2} < 30~\gev
\end{eqnarray*}
requirement in the VR and
\begin{eqnarray*}
30~\gev < \sqrt{ \left( m_{2b}^{\mathrm{lead}} - 123.7~\gev\right)^2 + \left( m_{2b}^{\mathrm{subl}} - 116.5~\gev\right)^2} < 45~\gev
\end{eqnarray*}
requirement in the SB. The $m_{2b}$ distributions of the leading versus subleading Higgs boson candidates for the non-resonant signal and the multijet
background are shown in Figure~\ref{fig:resolvedRegions}, together with the contours of the SR, VR and SB.

\begin{figure}
\begin{center}
\subfigure[]{\includegraphics[width=0.49\textwidth]{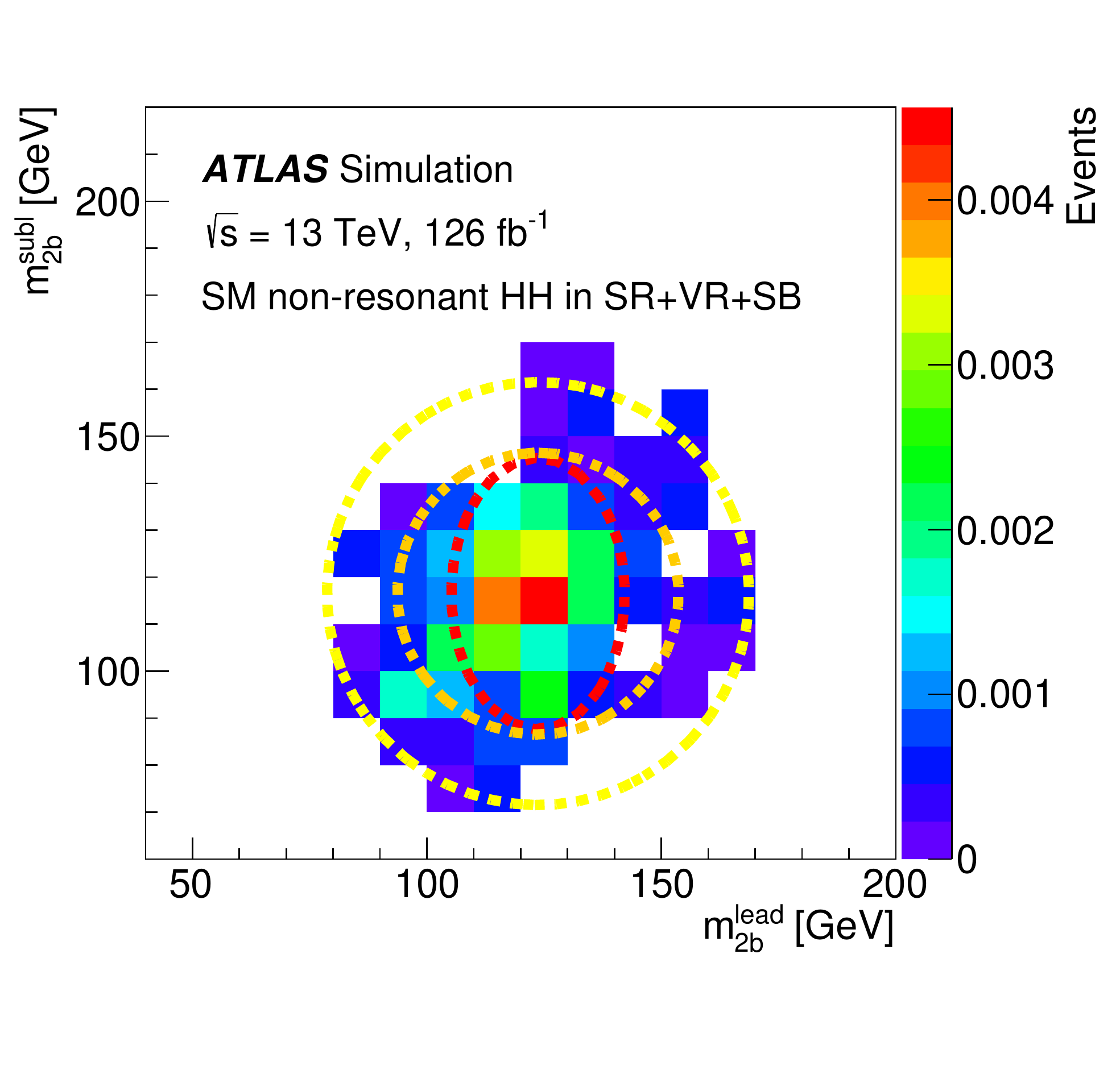}\label{fig:resolvedRegions:nores}}
\subfigure[]{\includegraphics[width=0.49\textwidth]{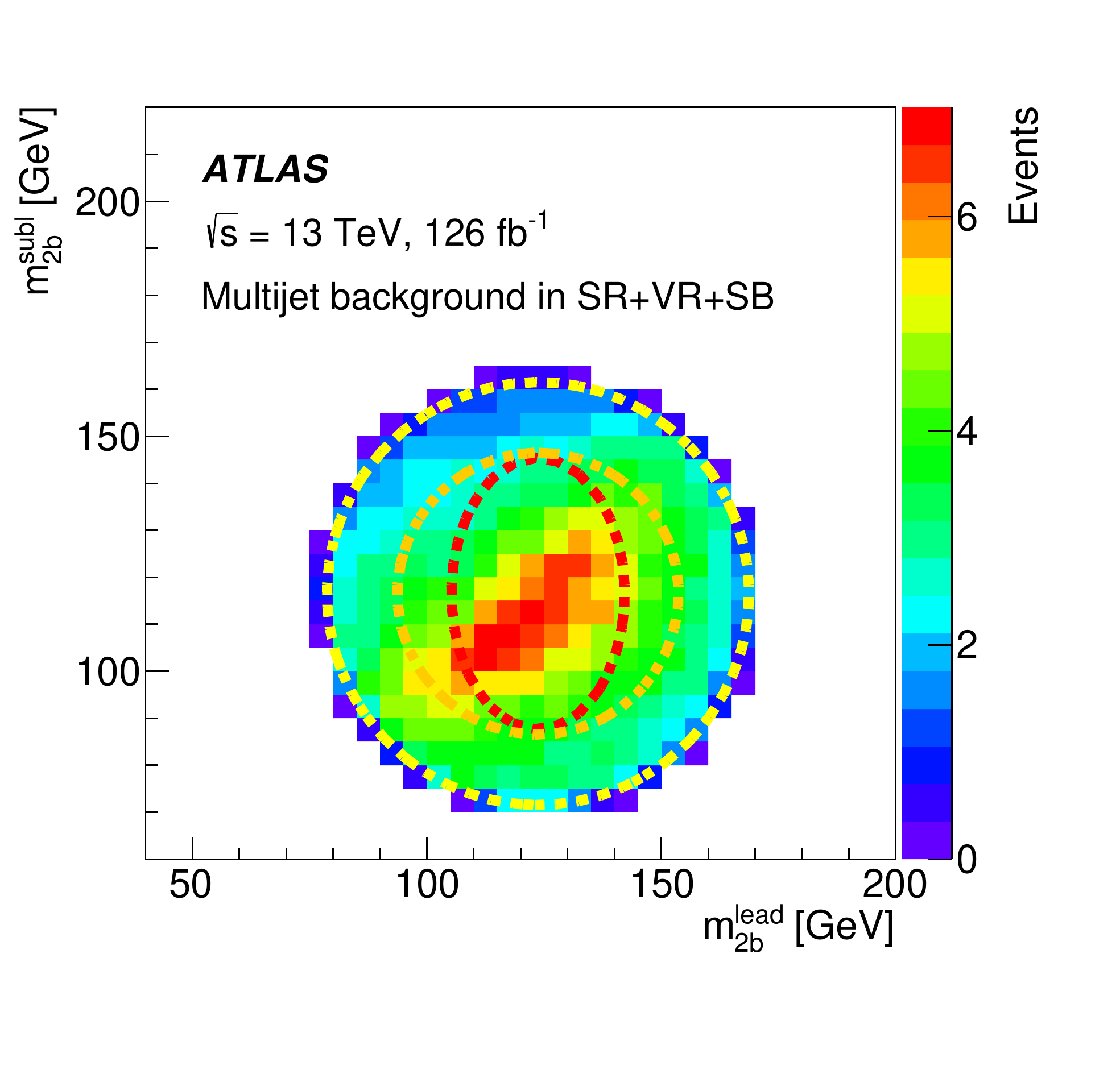}\label{fig:resolvedRegions:mj}}
\end{center}
\caption{Two-dimensional mass regions used in the analysis. The signal region is inside the inner (red) dashed curve, the validation region is outside the signal region and within the
intermediate (orange) circle, and the sideband is outside the validation region and within the outer (yellow) circle. The regions are shown for \subref{fig:resolvedRegions:nores}~simulated
events from the SM non-resonant $HH$ process and \subref{fig:resolvedRegions:mj}~the estimated multijet background.}
\label{fig:resolvedRegions}
\end{figure}

 
 
\section{Background estimation}
\label{sec:background}
 
After the event selection described in Section~\ref{sec:selection}, the background is dominated by multijet and \ttbar\ events.
The multijet events constitute about 95\% of the total background and are modelled using data. The remaining 5\% are \ttbar\
events, which are modelled using simulation. The normalisation of the all-hadronic \ttbar\ background is determined from data,
whereas the non-all-hadronic \ttbar\ background is normalised to the SM prediction. In the SM, the contribution of the $HH$ pairs
produced via ggF is small compared to other backgrounds and three times larger than for the VBF production. Thus
$HH$ production via ggF is treated as a background in this analysis and is fixed to the SM prediction. Other minor backgrounds
with contributions below 0.5\% are neglected. The background estimation technique is the same as in Ref.~\cite{EXOT-2016-31}.
 
\subsection{Multijet background}
\label{sec:multijet_background}
 
The data-driven multijet background estimation uses data events with lower $b$-jet multiplicity and reweights them to model
events with higher $b$-jet multiplicity. The multijet events are selected using the same trigger and selection requirements as those
used in the SR, except for the $b$-tagging requirement. In particular, the SR requires at least four $b$-jets (``four-tag sample'').
To derive a background estimate for this region, events with at least four central jets, but with only two of them $b$-tagged (``two-tag sample''),
are used. The events in the two-tag sample are reweighted by applying a product of two event weights. The first event weight corrects
for the additional $b$-tagged jet activity and the second event weight corrects for the kinematic differences caused by requiring
additional $b$-tagged jets. These differences can arise for a variety of reasons: the $b$-tagging efficiency varies as a function of jet \pt
and $\eta$; the various multijet processes contribute with different fractions in each sample; and the fraction of events accepted by each
trigger path changes. The reweighting is performed using one-dimensional distributions and is iterated until the weights converge to
stable values. Details of the reweighting procedure can be found in Ref.~\cite{EXOT-2016-31}. The weights are derived in the SB using
the procedure described above and validated in the VR.
 
\subsection{\ttbar\ background}
 
The shape of the \ttbar\ background is modelled using simulation. The \ttbar\ events are expected to contain two $b$-jets from the decay
of two top quarks and additional jets stemming from the hadronic $W$-boson decay or additional quarks or gluons produced together
with two top quarks. To reduce the statistical uncertainty, simulated \ttbar\ events in the two-tag region, corrected by the kinematic
weights derived for the multijet background, are also used. The procedure is validated in the SB, and good agreement is observed
between the corrected two-tag sample and the four-tag sample within the uncertainties. Samples for all-hadronic and non-all-hadronic
\ttbar\ decays are handled separately.

\subsection{Background normalisation}
 
The normalisations of multijet and all-hadronic \ttbar\ backgrounds are derived simultaneously by fitting the \Xwt\ distribution to data in
the SB. The \Xwt\ distribution differs for the two backgrounds: the region of $\Xwt < 0.75$ is enriched in all-hadronic \ttbar\ events
and the region $\Xwt > 0.75$ is enriched in multijet events. The normalisation of the non-all-hadronic \ttbar\ background is fixed to
its SM prediction in the fit due to its small contribution to the total yields. Two parameters are used in the normalisation fit: \mumj\ and
\muttbar. The \mumj\ parameter scales the multijet yield from the two-tag to the four-tag sideband region after the reweighting
described in Section~\ref{sec:multijet_background}. The \muttbar\ parameter corrects the normalisation of all-hadronic \ttbar\ yields
in the four-tag sideband region. The fit results are cross-checked in the VR, where the same results are obtained within statistical uncertainties.

 
 
\section{Systematic uncertainties}
\label{sec:systematics}
 
Background normalisation uncertainties are propagated from the fit, described in the previous section, which determines the multijet and all-hadronic \ttbar\ yields. The statistical
uncertainty of the multijet and all-hadronic \ttbar\ normalisation parameters is accounted for,
including their correlations. Two nuisance parameters are defined in the final fit described in Section~\ref{sec:results} by calculating two eigenvectors from the covariance matrix
of the normalisation fit. Furthermore, the normalisation of the multijet background
estimate is verified in the VR, where agreement with data is found; its statistical uncertainty in the VR is thus applied as a normalisation systematic uncertainty of the multijet background.
 
Two shape uncertainties for the multijet background modelling are evaluated using data from the VR. The multijet modelling uncertainty is related to the level of
agreement between the data and the background model in this region. To evaluate this uncertainty, the $\mfb$ distribution is split into low and high mass regions
at 400~\gev\ and a linear fit to the ratio of this distribution between data and the sum of all backgrounds is performed in both mass regions. Studies show that
the ratio, which appears to be constant, can be sufficiently well described by a straight-line fit, and that the change in the result is marginal for other splitting points. The $+1 \sigma$
variation of the slope of the fitted line and the $+1 \sigma$ variation of the fitted line with inverted slope are used as up and down variations of the uncertainty
in the background shape. The yield of the varied multijet templates is fixed to its nominal value. The varied templates are derived separately for the low- and high-$\mfb$ mass regions,
the corresponding systematics are treated as uncorrelated in the final fit.
The kinematic reweighting uncertainty is assessed by deriving an alternative multijet template using the same procedure as
in the nominal case, but using data from the VR. This difference between the multijet templates derived in the SB and the VR is symmetrised around
the nominal template, keeping the yield fixed to its nominal value.
 
The shape modelling uncertainty of the \ttbar\ background is evaluated by comparing the nominal \ttbar\ template derived using the two-tag
sample as described in Section~\ref{sec:background} and a \ttbar\ template derived using the four-tag sample after the basic preselection
requirements defined in Section~\ref{sec:selection}. A straight-line fit to the ratio of the two-tag to four-tag $\mfb$ templates is performed. With a procedure similar to
the one applied for the multijet shape uncertainties, the up and down variations around the nominal \ttbar\ template are extracted using a straight-line fit to the $+1 \sigma$
variation of the slope of the fitted line and the $+1 \sigma$ variation of the fitted line with inverted slope, while keeping the yield fixed to its nominal value. This
uncertainty is derived separately for the non-all-hadronic (using MC) and all-hadronic (using data) \ttbar\ samples and is not correlated in the fit between these two samples.
 
Theoretical uncertainties in the ggF background yield are evaluated by varying the renormalisation and factorisation scales and from the uncertainty associated with the choice of PDF set.
The resulting variation of the expected ggF background yield is about 10\%. When considering the same sources of theoretical uncertainty for the VBF $HH$ signal,
its acceptance times efficiency varies by 3\%.
 
The experimental uncertainties listed below affect only MC samples. The uncertainties in the jet energy
resolution and scale are evaluated at $\sqrt{s} = 13$~\TeV\ using in situ measurement techniques described in Ref.~\cite{PERF-2016-04}. The sources of uncertainty in these measurements are treated as
fully correlated between the \pT\ and mass scales. The resolution uncertainty is evaluated in measurements documented in
Ref.~\cite{PERF-2011-04} and is assessed by applying an additional smearing to these observables.
The flavour tagging efficiency and its uncertainty for $b$- and $c$-jets is estimated in $t\bar{t}$ events, while the light-jet misidentification rate
and uncertainty is determined using dijet events~\cite{FTAG-2018-01,ATLAS-CONF-2018-001,ATLAS-CONF-2018-006}. In addition, an
uncertainty in the $b$-jet trigger efficiency is derived from the per-jet online $b$-tagging measurements~\cite{EXOT-2016-33}. The uncertainty in the integrated luminosity is 1.7\%~\cite{ATLAS-CONF-2019-021}, obtained using the LUCID-2 detector~\cite{LUCID2} for the primary luminosity measurements.
 

 
\section{Results}
\label{sec:results}
 
Following the statistical procedures outlined in Ref.~\cite{HIGG-2012-27}, a test statistic based on the profile likelihood ratio~\cite{Cowan:2010js} is
used to test hypothesised values $\sigma_\text{VBF}$ of the cross-section of the signal model in units of fb. This test statistic extracts
the information about the signal cross-section from a likelihood fit to the data. The likelihood function includes all parameters which describe the
systematic uncertainties and their correlations discussed in Section~\ref{sec:systematics}. As no significant excess over the background prediction is
observed, exclusion limits are computed using the asymptotic formula~\cite{Cowan:2010js}. The exclusion limits are based on the CL$_\text{s}$
method~\cite{Read:2002hq}, where a value of $\sigma_\text{VBF}$ is regarded as excluded at the 95\% confidence level (CL) when CL$_\text{s}$
is smaller than 5\%. The accuracy of the asymptotic approximation is verified with sampling distributions generated using pseudo-experiments.
 
The mass of the four selected $b$-jets, \mfb, in the SR is used as the final discriminant for limit setting. Figure~\ref{fig:massdist} shows the distribution
of data and the SM background after the background-only fit in the SR and VR. In addition, the signal prediction for the narrow-width resonance hypothesis
with $m_X = 800$~\GeV\ and the non-resonant signal at $\ktv = 3$ are shown in the SR.
 
\begin{figure}
\begin{center}
\subfigure[]{\includegraphics[width=0.49\textwidth]{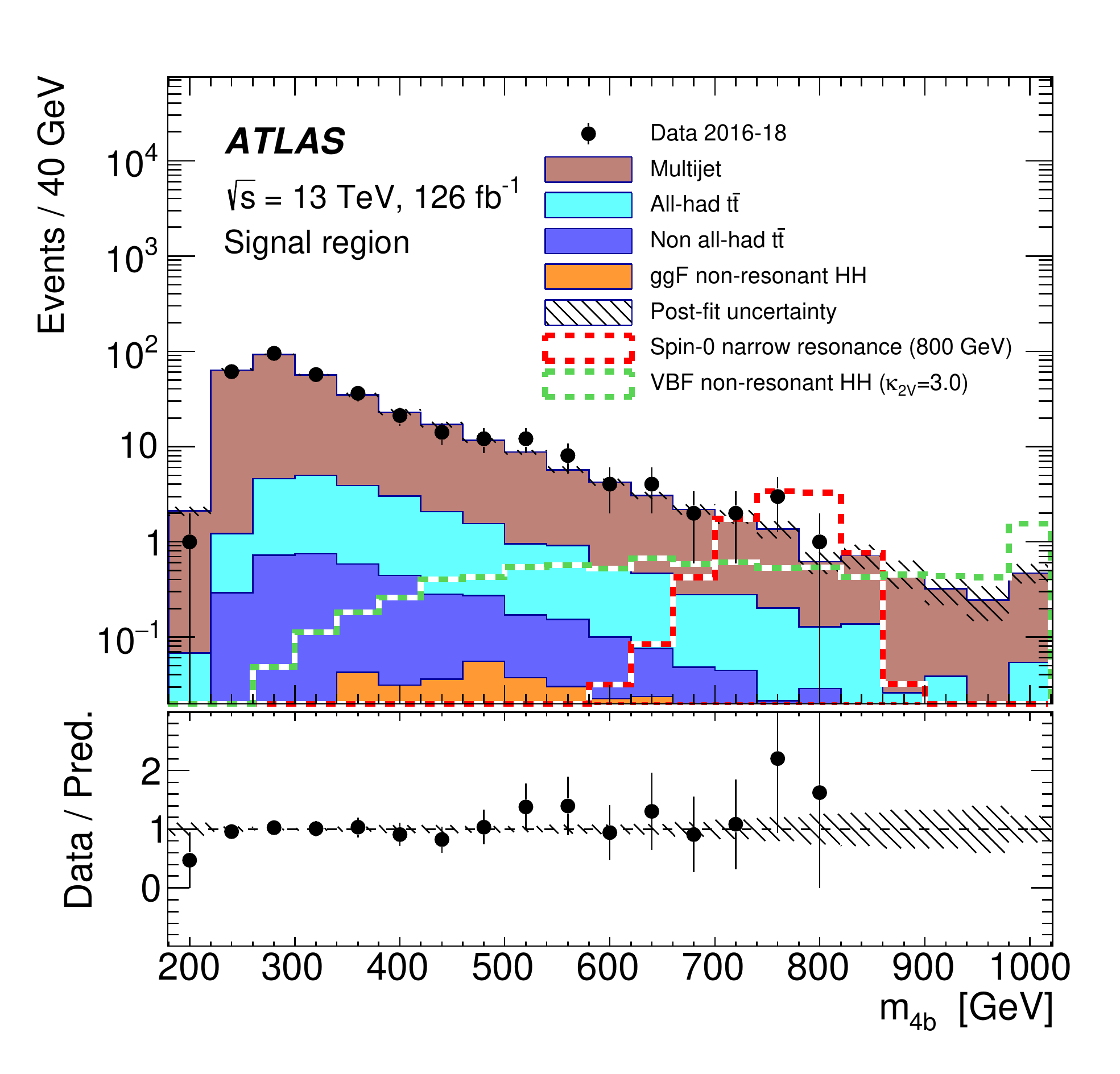}\label{fig:SRmassdist}}
\subfigure[]{\includegraphics[width=0.49\textwidth]{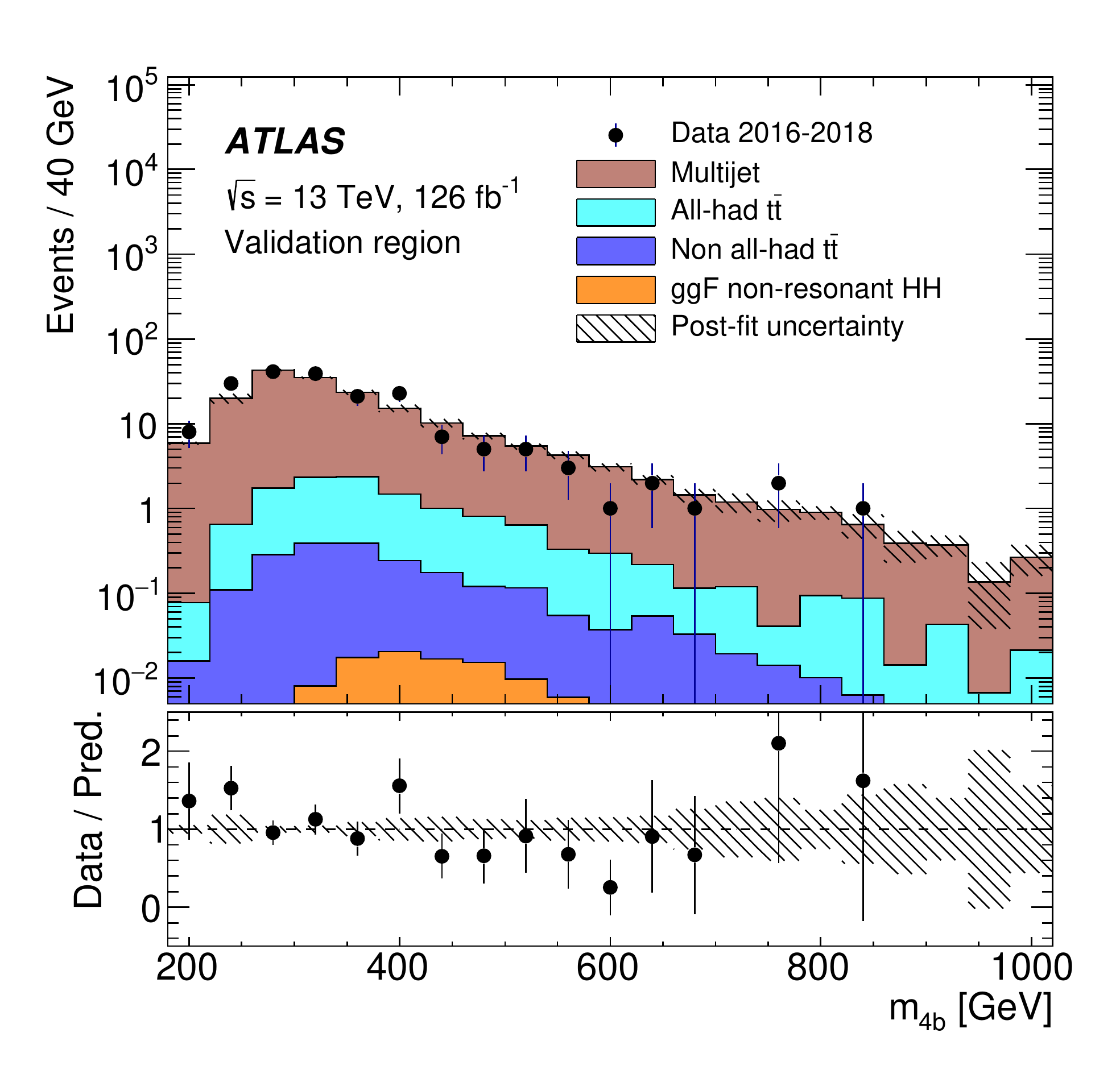}\label{fig:addmassdist:VR}}
\end{center}
\caption{Post-fit mass distribution of the $HH$ candidates in the~\subref{fig:SRmassdist} signal and~\subref{fig:addmassdist:VR}
validation regions. The expected background is shown after the profile-likelihood fit to data with the background-only hypothesis; the narrow-width resonant
signal at 800~\gev\ and the non-resonant signal at $\ktv=3$ are overlaid in the signal region, both normalised to the corresponding observed upper limits on
the cross-section. The lower panel shows the ratio of the observed data to the estimated SM background. The distribution of events is shown per mass interval
corresponding to the bin width of 40~\gev, while the overflow events are included in the last bin.}
\label{fig:massdist}
\end{figure}

Upper limits on the cross-section are set for all tested models. Figure~\ref{fig:CombinedLimits} shows the 95\% CL upper limits for resonant $HH$ production via VBF
as a function of the resonance mass $m_X$ for narrow- and broad-width resonance hypotheses. The significance of the excess over the background-only prediction is quantified using the local $p_0$-value, defined as the probability of the
background-only model to produce a signal-like fluctuation at least as large as that observed in the data. The most extreme $p_0$-value corresponds
to a local significance of $1.5$ standard deviations at 550~\gev.

\begin{figure}
\begin{center}
\subfigure[]{\includegraphics[width=0.49\textwidth]{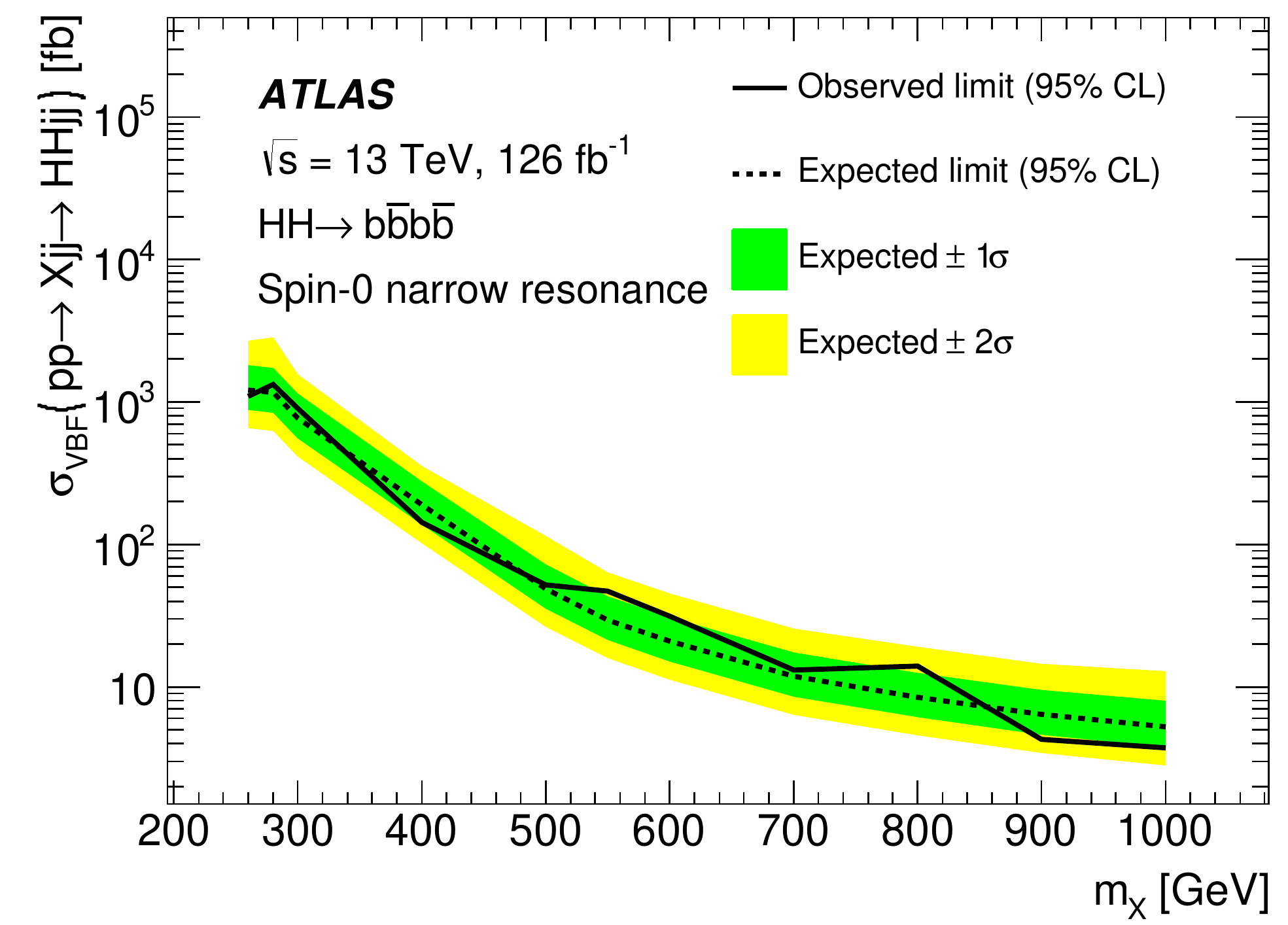}\label{fig:CombinedLimits:narrow}}
\subfigure[]{\includegraphics[width=0.49\textwidth]{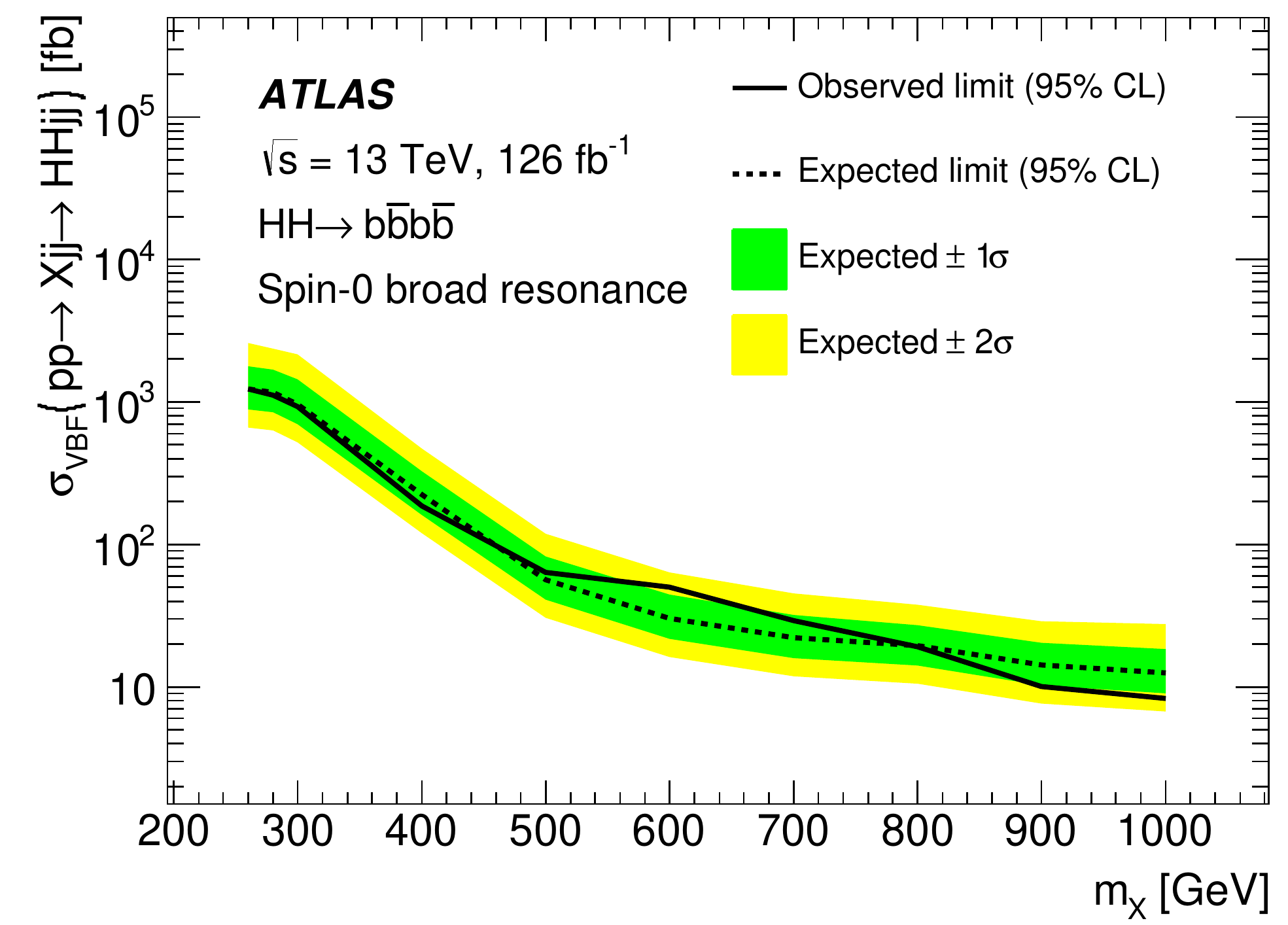}\label{fig:CombinedLimits:broad}}
\caption{Observed and expected 95\% CL upper limits on the production cross-section for resonant $HH$ production via VBF as a function of the
mass $m_X$. The \subref{fig:CombinedLimits:narrow} narrow- and \subref{fig:CombinedLimits:broad} broad-width resonance hypotheses are presented.}
\label{fig:CombinedLimits}
\end{center}
\end{figure}
 
The expected and observed limits on SM non-resonant $HH$ production via VBF are given in Table~\ref{tab:smrwMhhlims}. Limits are also calculated as a function
of \ktv, as presented in Figure~\ref{fig:c2vLimits}. The observed excluded region corresponds to $\ktv < -0.43$ and $\ktv > 2.56$, while the expected exclusion is
$\ktv < -0.55$ and $\ktv > 2.72$. For \ktv\ values deviating from the SM prediction, growing non-cancellation effects result in a harder $m_{HH}$ spectrum, and
thereby higher-\pt $b$-jets, which in turn lead to increased signal acceptance times efficiency as shown in Figure~\ref{fig:resolvedEfficiency}. This search is therefore
not sensitive to the region close to the SM prediction, corresponding to $\ktv = 1$ .
 
\begin{table}[htp]
\caption{Upper limits at 95\% CL for SM non-resonant $HH$ production via VBF in fb (first row) and normalised to its SM expectation,
$\sigma^\text{SM}_\text{VBF}$ (second row). Uncertainties related to the branching ratio of the $H \to b \bar{b}$ decay are not considered.}
\begin{center}
\begin{tabular}{|c|c|ccccc|}
\hline
&Observed & $-2\sigma$ & $-1\sigma$ & Expected & $+1\sigma$ & $+2\sigma$ \\
\hline
$\sigma_\text{VBF}$ [fb] & 1460 & 510 & 690 & 950 & 1330 & 1780 \\ 
$\sigma_\text{VBF}/\sigma^\text{SM}_\text{VBF}$& ~~840 & 290 & 400 & 550 & ~~770 & 1030 \\ 
 
\hline
\end{tabular}
\end{center}
\label{tab:smrwMhhlims}
\end{table}
 
\begin{figure}
\begin{center}
\includegraphics[width=0.65\textwidth]{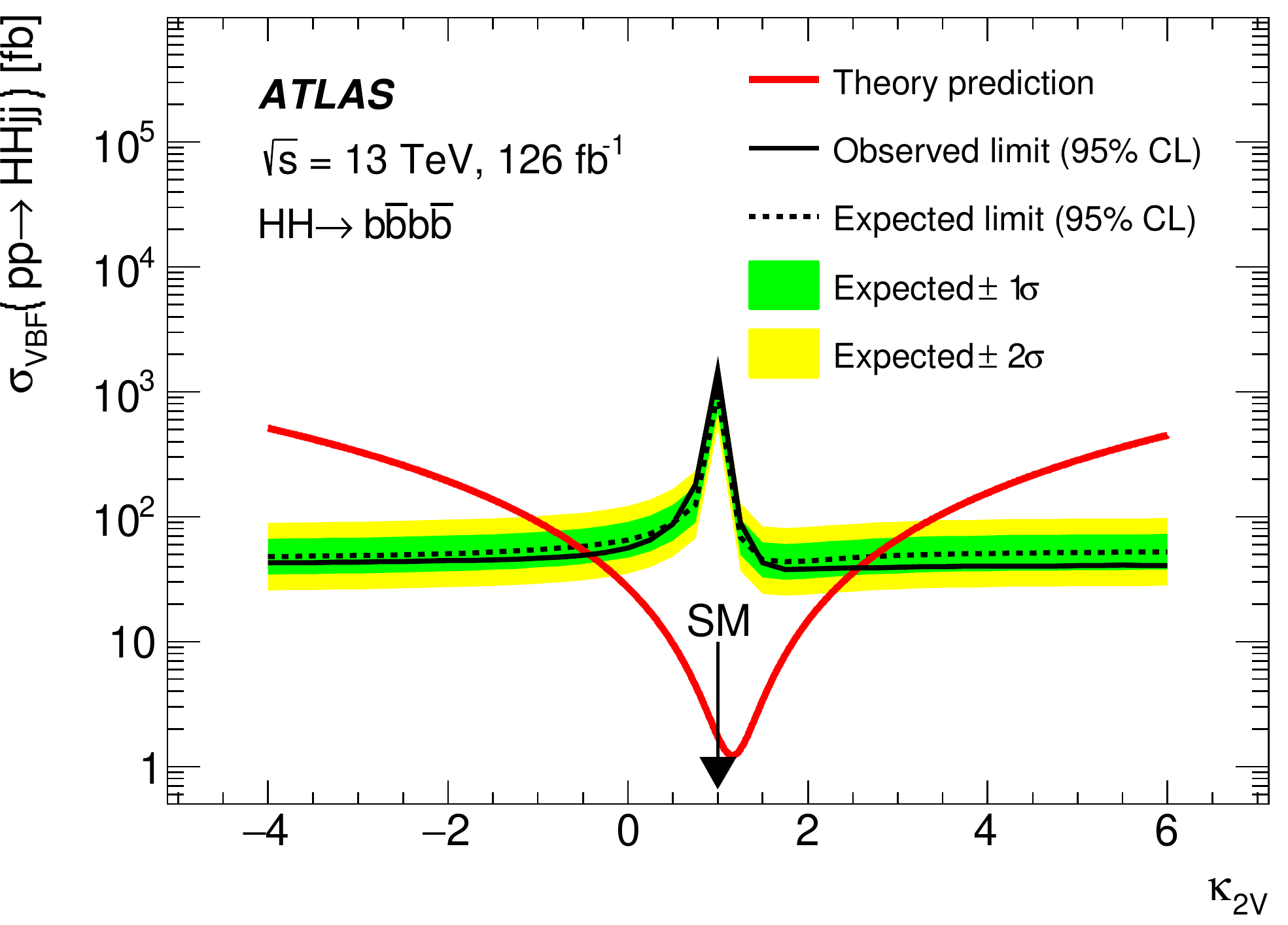}
\caption{Observed and expected 95\% CL upper limits on the production cross-section for non-resonant $HH$ production via VBF as a
function of the di-vector-boson--di-Higgs-boson coupling modifier \ktv. The theory prediction of the cross-section
as a function of \ktv\ is also shown.  More details on the predicted cross-section can be found in Section~\ref{sec:samples}. }
\label{fig:c2vLimits}
\end{center}
\end{figure}
 
Table~\ref{tab:npranking} summarises the relative impact of the uncertainties on the best-fit signal cross-section for two different narrow-width resonance
production hypotheses, with masses equal to 300~\gev\ and 800~\gev. Only major sources of systematic uncertainty are
quoted along with the impact of the statistical uncertainty. The uncertainties of similar nature are grouped into unique categories and the fit is performed independently
for the two hypothesised signals. The systematic uncertainties related to the multijet background estimate have the largest impact on the result.
 
\begin{table}
\caption{Dominant relative uncertainties in the best-fit signal cross-section $\sigma^\text{best fit}_\text{VBF}(pp\rightarrow Xjj \rightarrow HHjj)$ of hypothesised resonant $HH$ signal production.
The leading sources of systematic uncertainty, the total systematic uncertainty and the data statistical uncertainty are provided. Two mass points are selected: $m_X = 300$~\gev\ with the
best-fit cross-section of $140$~fb and $m_X = 800$~\gev\ with $4.7$~fb, which correspond to the low and high mass regions.
The groups of uncertainties do not add up in quadrature to the total uncertainty, because only the dominant uncertainties are shown and also due to correlations between the uncertainties.}
\centering
{
\scriptsize
\begin{tabular}{|lr|lr|}
\hline
Source & $m_X =$ 300~\gev & Source & $m_X =$ 800~\gev \\
\hline
Multijet normalisation  & 46\% & Multijet modelling & 44\% \\
Jet energy resolution& 26\% & Jet energy resolution & 23\% \\
Multijet modelling  & 18\% & Jet energy scale & 19\% \\
Multijet kinematic reweighting  & 17\% & Multijet kinematic reweighting & 9\%\\
\ttbar\ modelling & 11\% & Multijet normalisation & 7\%\\
Jet energy scale & 10\% & \ttbar modelling & 6\%\\
\hline
Total systematic uncertainty  & 64\% & Total systematic uncertainty  & 57\% \\
Statistical uncertainty       & 77\% & Statistical uncertainty & 82\% \\
\hline
\end{tabular}
\label{tab:npranking}
}
\end{table}

 
\FloatBarrier
\section{Conclusion}
 
A search for both resonant and non-resonant production of pairs of Standard Model Higgs bosons via
vector-boson fusion has been carried out in the \bbbb\ channel. The analysed data were collected from
$\sqrt{s} =$ 13~\tev\ proton--proton collisions by the ATLAS detector at the LHC in 2016--2018 and
correspond to an integrated luminosity of \lumi~\ifb. Results for resonant $HH$ production are reported in the mass range
\mlow--\mhigh~\gev. The largest deviation from the background-only hypothesis is observed
at 550~\gev\ with a local significance of 1.5 standard deviations. Upper limits on the production cross-section are set for narrow-
and broad-width scalar resonances at 95\% CL. Limits are also set on the cross-section for non-resonant $HH$ production, and as
a function of the di-vector-boson--di-Higgs-boson coupling modifier, \ktv. The observed 95\% CL upper limit on the SM non-resonant
$HH$ production cross-section is 1460~fb, compatible with the expected limit at a level below two standard deviations. The observed
excluded region corresponds to $\ktv < -0.43$ and $\ktv > 2.56$, while the expected exclusion is $\ktv < -0.55$ and $\ktv >$ 2.72.

\section*{Acknowledgements}
 

We thank CERN for the very successful operation of the LHC, as well as the
support staff from our institutions without whom ATLAS could not be
operated efficiently.
 
We acknowledge the support of ANPCyT, Argentina; YerPhI, Armenia; ARC, Australia; BMWFW and FWF, Austria; ANAS, Azerbaijan; SSTC, Belarus; CNPq and FAPESP, Brazil; NSERC, NRC and CFI, Canada; CERN; ANID, Chile; CAS, MOST and NSFC, China; COLCIENCIAS, Colombia; MSMT CR, MPO CR and VSC CR, Czech Republic; DNRF and DNSRC, Denmark; IN2P3-CNRS and CEA-DRF/IRFU, France; SRNSFG, Georgia; BMBF, HGF and MPG, Germany; GSRT, Greece; RGC and Hong Kong SAR, China; ISF and Benoziyo Center, Israel; INFN, Italy; MEXT and JSPS, Japan; CNRST, Morocco; NWO, Netherlands; RCN, Norway; MNiSW and NCN, Poland; FCT, Portugal; MNE/IFA, Romania; JINR; MES of Russia and NRC KI, Russian Federation; MESTD, Serbia; MSSR, Slovakia; ARRS and MIZ\v{S}, Slovenia; DST/NRF, South Africa; MICINN, Spain; SRC and Wallenberg Foundation, Sweden; SERI, SNSF and Cantons of Bern and Geneva, Switzerland; MOST, Taiwan; TAEK, Turkey; STFC, United Kingdom; DOE and NSF, United States of America. In addition, individual groups and members have received support from BCKDF, CANARIE, Compute Canada, CRC and IVADO, Canada; Beijing Municipal Science \& Technology Commission, China; COST, ERC, ERDF, Horizon 2020 and Marie Sk{\l}odowska-Curie Actions, European Union; Investissements d'Avenir Labex, Investissements d'Avenir Idex and ANR, France; DFG and AvH Foundation, Germany; Herakleitos, Thales and Aristeia programmes co-financed by EU-ESF and the Greek NSRF, Greece; BSF-NSF and GIF, Israel; La Caixa Banking Foundation, CERCA Programme Generalitat de Catalunya and PROMETEO and GenT Programmes Generalitat Valenciana, Spain; G\"{o}ran Gustafssons Stiftelse, Sweden; The Royal Society and Leverhulme Trust, United Kingdom.
 
The crucial computing support from all WLCG partners is acknowledged gratefully, in particular from CERN, the ATLAS Tier-1 facilities at TRIUMF (Canada), NDGF (Denmark, Norway, Sweden), CC-IN2P3 (France), KIT/GridKA (Germany), INFN-CNAF (Italy), NL-T1 (Netherlands), PIC (Spain), ASGC (Taiwan), RAL (UK) and BNL (USA), the Tier-2 facilities worldwide and large non-WLCG resource providers. Major contributors of computing resources are listed in Ref.~\cite{ATL-SOFT-PUB-2020-001}.
 
 
\clearpage

\printbibliography
 
\clearpage
 
\begin{flushleft}
\hypersetup{urlcolor=black}
{\Large The ATLAS Collaboration}

\bigskip

\AtlasOrcid[0000-0002-6665-4934]{G.~Aad}$^\textrm{\scriptsize 102}$,    
\AtlasOrcid[0000-0002-5888-2734]{B.~Abbott}$^\textrm{\scriptsize 129}$,    
\AtlasOrcid{D.C.~Abbott}$^\textrm{\scriptsize 103}$,    
\AtlasOrcid[0000-0002-2788-3822]{A.~Abed~Abud}$^\textrm{\scriptsize 71a,71b}$,    
\AtlasOrcid[0000-0002-1002-1652]{K.~Abeling}$^\textrm{\scriptsize 53}$,    
\AtlasOrcid[0000-0002-2987-4006]{D.K.~Abhayasinghe}$^\textrm{\scriptsize 94}$,    
\AtlasOrcid[0000-0002-8496-9294]{S.H.~Abidi}$^\textrm{\scriptsize 167}$,    
\AtlasOrcid[0000-0002-8279-9324]{O.S.~AbouZeid}$^\textrm{\scriptsize 40}$,    
\AtlasOrcid{N.L.~Abraham}$^\textrm{\scriptsize 156}$,    
\AtlasOrcid[0000-0001-5329-6640]{H.~Abramowicz}$^\textrm{\scriptsize 161}$
% ...
\bigskip
\\

% ...
$^{40}$Niels Bohr Institute, University of Copenhagen, Copenhagen; Denmark.\\
% ...
$^{53}$II. Physikalisches Institut, Georg-August-Universit\"{a}t G\"ottingen, G\"ottingen; Germany.\\
% ...
$^{71}$$^{(a)}$INFN Sezione di Pavia;$^{(b)}$Dipartimento di Fisica, Universit\`a di Pavia, Pavia; Italy.\\
% ...
$^{94}$Department of Physics, Royal Holloway University of London, Egham; United Kingdom.\\
% ...
$^{102}$CPPM, Aix-Marseille Universit\'e, CNRS/IN2P3, Marseille; France.\\
$^{103}$Department of Physics, University of Massachusetts, Amherst MA; United States of America.\\
% ...
$^{129}$Homer L. Dodge Department of Physics and Astronomy, University of Oklahoma, Norman OK; United States of America.\\
% ...
$^{156}$Department of Physics and Astronomy, University of Sussex, Brighton; United Kingdom.\\
% ...
$^{161}$Raymond and Beverly Sackler School of Physics and Astronomy, Tel Aviv University, Tel Aviv; Israel.\\
% ...
$^{167}$Department of Physics, University of Toronto, Toronto ON; Canada.\\
% ...
$^{a}$ Also at Borough of Manhattan Community College, City University of New York, New York NY; United States of America.\\
$^{b}$ Also at CERN, Geneva; Switzerland.\\
% ...

\end{flushleft}


\end{document}